\documentclass[10pt,a4paper]{article}
\usepackage[latin1]{inputenc}
\usepackage{amsmath}
\usepackage{amsfonts}
\usepackage{amssymb}
\usepackage{geometry}
\usepackage{mathenv}
\usepackage{graphicx}
\usepackage{color}
\usepackage[affil-it]{authblk}
\usepackage{multicol}
\usepackage{slashbox}
\usepackage[english]{babel}
\title{A Green correlation approach for passive identification - Application to acoustic and solid waves with a viscous damping model}
\author{M. Carmona$^1$, OJJ. Michel$^2$, J-L. Lacoume$^{1,2}$, N. Sprynski$^1$ and B. Nicolas$^2$}
\affil{$^{1}$CEA-LETI, MINATEC-Campus, 38054 Grenoble, France \\ 
$^{2}$Gipsa-lab, BP 46 F- 38402 Grenoble Cedex, France}

\begin{document}

\maketitle
\abstract{This paper presents a new approach on passive identification of elastic propagation media. As passive identification relies upon noise field correlation, an original and perhaps more natural approach is developed which consists in considering the Green correlation. It is shown that Green correlation contains all medium parameters and provides an appealing alternative to the classical Green function estimation through a Ward identity. Our formalism allows to extend classical scalar passive identification models to vectorial ones and to take into account realistic dissipation models. This approach is applied to acoustic and solid waves with viscous damping. }  

\section{Introduction}
\label{intro}

\indent \indent Passive identification of a propagation medium consists in retrieving medium parameters, by using uncontrolled noise fluctuations \textcolor{blue}{\cite{Snieder}}. Such an idea has long been pursued in acoustics (\textcolor{blue}{\cite{Lobkis}}, \textcolor{blue}{\cite{Snieder}}) and seismology (\textcolor{blue}{\cite{Campillo}}, \textcolor{blue}{\cite{Wapenaar}}) and gave rise to numerous applications and experimental validations (\textcolor{blue}{\cite{Sabra}}, \textcolor{blue}{\cite{Miyazawa}}, \textcolor{blue}{\cite{Gouedard}}). \\
\indent Preceding studies (\textcolor{blue}{\cite{Lobkis}}, \textcolor{blue}{\cite{Campillo}}, \textcolor{blue}{\cite{Sabra}}, \textcolor{blue}{\cite{Miyazawa}}, \textcolor{blue}{\cite{Gouedard}}) rely upon the estimation of the Green function of the medium. Such estimation is made possible by exploiting a Ward identity \textcolor{blue}{\cite{Weaver}}, which relates the noise correlation function to the Green function (\textcolor{blue}{\cite{Lobkis}}, \textcolor{blue}{\cite{Snieder}}, \textcolor{blue}{\cite{ColinDeVerdiere}}). The fundamental role of dissipation in Ward identity  was outlined in \textcolor{blue}{\cite{Gouedard}}, were dissipation is assumed to be constant; however, a constant dissipation model is hardly acceptable from a physical point of view (\textcolor{blue}{\cite{Landau}}, \textcolor{blue}{\cite{Royer}}), and needs to be further  discussed. \\

	The contribution of the paper is twofold. Firstly, an alternative approach to passive identification is proposed, which is based upon the Green correlation function. Green correlation is introduced as the correlation of a propagated white noise field \textcolor{blue}{\cite{Lacoume}}. We show that Green correlation contains all physical parameters that need to be identified for a complete characterization of the propagation medium. The motivation for introducing Green correlation comes from classical system identification theory, where noise based identification relies strongly on the transformation of second order statistics through linear systems \textcolor{blue}{\cite{Ljung}}.\\
\indent Secondly, we highlight the role of Green correlation in the framework of elastic waves with viscous damping. Green function and Green correlation are explicitly formulated in the time and space Fourier formalism. Ward identities are also developed. Furthermore, for low viscous damping and far-field assumptions, a new formulation of the Ward identity involving third order time derivative, is derived. \\

	The organization of the paper is the following. In \textcolor{blue}{\textbf{section 2}}, we introduce passive identification through a linear system approach. We recall the definition of the Green function and its role in medium identification. Cross-correlation of random fields is recalled, and white noise notion is introduced. Then, we define the Green correlation, we show its role in passive identification and we discuss its relation with the Green function through Ward identities. \\
\indent In \textcolor{blue}{\textbf{section 3}}, solid waves equation with viscous damping is presented. Dispersion matrix and relation of dispersion are introduced. Acoustic waves propagation is seen as a particular case of solid propagation study. \\
\indent Solid and acoustic Green functions are computed in \textcolor{blue}{\textbf{section 4}}. We highlight their role in solid and acoustic medium identification, respectively. Approximation of far-field and low attenuation is also considered. This particular, but realistic, case allows to derive explicit expressions.\\
\indent Solid and acoustic Green correlation are computed in \textcolor{blue}{\textbf{section 5}}. The role of Green correlation in passive identification is emphasized in those practical cases. Ward identities are derived and compared to existing ones for a constant damping model.  

\section{Green function and Green correlation of a linear propagation medium} 
\label{sec:2}

\indent \indent	In this section, we adopt a linear system approach to describe medium identification. We introduce the Green function and we relate it to the classical impulsional response of linear systems. White noise is introduced and its existence is discussed. From white noise, we define the Green correlation and we highlight its natural role in passive identification. 

\subsection{Medium and fields: a system approach.}
\label{sec:21}

\indent \indent	We denote by $\underline{\mathbf{u}}(t,\underline{x})$ the value of field $\underline{\mathbf{u}}$ at time $t$ and position $\underline{x}$. When $\underline{\mathbf{u}}$ has one component it is said scalar, otherwise it is said vectorial. \\
	
	A propagation medium can be seen as a system where the source field $\underline{\mathbf{f}}$ is the input and the generated field $\underline{\mathbf{u}}$ is the output of the system. Those two fields are related by a relation of the following type:  
\begin{eqnarray}
\underline{\mathbf{u}}=\mathfrak{G}(\underline{\mathbf{f}})
\label{u=G(f)}
\end{eqnarray}
where operator $\mathfrak{G}$ is a representation of medium properties: physical laws, boundary and initial conditions. The medium is said linear when $\mathfrak{G}$ is a linear operator. Only linear medium are considered in this paper. \\

	With that formalism, medium identification can be seen as system identification where parameters are physical, like attenuation or propagation speed, and geometrical, like distance or orientation between sensors. 

\subsection{Green function of a linear medium.}
\label{sec:22}

\indent \indent	A linear medium $X$ satisfies the superposition theorem \textit{i.e.} the value $\underline{\mathbf{u}}(t,\underline{x})$ of the generated field can be seen as the superposition of all contributions of elementary sources $\underline{\mathbf{f}}(t',\underline{x}')dt'd\underline{x}'$ emitted at time $t'$ during $dt'$ period in the volume centered in $\underline{x}'$ and of dimensions $d\underline{x}'$, for all times $t'$ and points $\underline{x}'$. Mathematically, this can be written $\underline{\mathbf{u}}(t,\underline{x})=\int_{\mathbb{R} \times X} \underline{\underline{\mathbf{G}}}(t,\underline{x},t',\underline{x}')\underline{\mathbf{f}}(t',\underline{x}')dt'd\underline{x}'$ and simplified by introducing the generalised convolution $\otimes_{T,S}$ as:
\begin{eqnarray}
\underline{\mathbf{u}}=\underline{\underline{\mathbf{G}}} \otimes_{T,S} \underline{\mathbf{f}}
\label{u=Gf}
\end{eqnarray}	
$\underline{\underline{\mathbf{G}}}$ is called the Green function of the medium as it is the kernel of the operator $\mathfrak{G}$ appearing in equation \textcolor{blue}{(\ref{u=G(f)})}. All medium mechanical parameters are contained in its expression. This highlights the importance of retrieving the Green function in medium identification. \\

	Physically, the $i$-th column of $\underline{\underline{\mathbf{G}}}$ corresponds to the medium response of a spatio-temporal impulsion directed by the $i$-th axis of the reference. According to this interpretation, $\underline{\underline{\mathbf{G}}}$ is sometimes called "impulsional response" of the medium in reference to the classical impulsional response of a linear system \textcolor{blue}{\cite{Ljung}}. Then, active identification consists in emitting spatio-temporal impulsions to retrieve the Green function and then to estimate model parameters \textcolor{blue}{\cite{Ljung}}.  \\
	
	We consider only time-shift invariant media. This property implies that $\underline{\underline{\mathbf{G}}}$ depends only on the times difference appearing in its parameters \textit{i.e.} we can do the following substitution: $\underline{\underline{\mathbf{G}}}(t,\underline{x},t',\underline{x}') \leftrightarrow \underline{\underline{\mathbf{G}}}(t-t',\underline{x},\underline{x}')$. 
	
\subsection{Cross-correlation of stochastic fields.}
\label{sec:23}	

\indent \indent In passive identification, source fields are not controlled. The principle relies on recording noise sources and using their statistical properties to retrieve medium parameters. With stochastic source fields, the analysis has to be performed from the cross-correlation of the generated field $\underline{\mathbf{u}}$ defined as:
\begin{eqnarray}
\underline{\underline{\mathbf{C}}}_{\; \underline{\mathbf{u}}}(t,\underline{x},t',\underline{x}')  :=  \mathrm{E} \left[ \underline{\mathbf{u}}(t,\underline{x}) \underline{\mathbf{u}}(t',\underline{x}')^{T}  \right]
\label{Cu}
\end{eqnarray} 	
where $\mathrm{E}$ and $T$ are the expectation operator and the transposition operator, respectively. \\

	We consider only stationary fields, this assumption is not a strong constraint in practice. In that case, cross-correlation depends only on the times difference appearing in its parameters \textit{i.e.} we can do the following substitution: $\underline{\underline{\mathbf{C}}}(t,\underline{x},t',\underline{x}') \leftrightarrow \underline{\underline{\mathbf{C}}}(t-t',\underline{x},\underline{x}')$. Furthermore, when fields are also considered ergodic in time, cross-correlation can be computed using the following formula:
\begin{eqnarray}
\underline{\underline{\mathbf{C}}}_{\; \underline{\mathbf{u}}}(t,\underline{x},\underline{x}') = \displaystyle{\lim_{\tau \to \infty}}\displaystyle{\int_0^{\tau}} \underline{\mathbf{u}}(t+t',\underline{x})\underline{\mathbf{u}}(t',\underline{x}')^T dt'
\label{CuE}
\end{eqnarray} 	
This formula is fundamental to approximate the cross-correlation of a field recorded by an array of sensors. 	
	
\subsection{White noise.}
\label{sec:24}		

\indent \indent By definition, a white noise is a field which value of a given time, position and direction is uncorrelated to any other value taken at all other times, positions and directions. Mathematically, the cross-correlation of a such field $\underline{\mathbf{f}}$ is a spatio-temporal isotropic impulsion: 
\begin{eqnarray}
\underline{\underline{\mathbf{C}}}_{\; \underline{\mathbf{f}}}(t,\underline{x},t',\underline{x}')= \delta (t,t')\delta(\underline{x},\underline{x}') \underline{\underline{\mathbf{I}}}_{\, p} 
\label{WN}
\end{eqnarray}
where $p$ is the number of components of $\underline{\mathbf{f}}$, $\underline{\underline{\mathbf{I}}}_{\, p}$ is the $p \times p$ identity matrix and $\delta$ is the Dirac distribution. \\

	A white noise has no physical reality in a sense that it has an infinite power. However, in practice the temporal whiteness is only needed in a limited frequency band. This latter is defined by the used instrumentation. The classical approach \textcolor{blue}{\cite{Campillo2}} to justify that ambient noise converges to a source with a spatio-isotropic whiteness consists to see the medium as a chaotic dynamical system. Then, according to equipartition theorem, it exists a time after which a coherent source snared in the medium becomes spatially and isotropically white. This time, called mixing time, depends on the frequency band, medium geometry and heterogeneity. With those considerations, cross-correlation of ambient noise is stacked during a sufficient long time in order to obtain a contribution of an approximated spatio-isotropic noise \textcolor{blue}{\cite{Campillo}}, \textcolor{blue}{\cite{Gouedard}}. 
	
\subsection{Green correlation.}
\label{sec:25}		

\indent \indent	As the Green function is the field generated by a spatio-temporal isotropic source \textit{i.e.} $\underline{\mathbf{f}}(t,\underline{x})=\delta(t,t')\delta(\underline{x},\underline{x}')\underline{\underline{\mathbf{I}}}_{\, p}$, we define by analogy the Green correlation $\underline{\underline{\mathbf{C}}}$ by the cross-correlation of a field generated by a white noise source. This function, introduced in \textcolor{blue}{\cite{Lacoume}}, plays by definition a fundamental role in passive identification. \\
	
	We can precise Green correlation expression using definition \textcolor{blue}{(\ref{Cu})} and equation \textcolor{blue}{(\ref{u=Gf})}. Indeed, for every generated field $\underline{\mathbf{u}}$, $\underline{\underline{\mathbf{C}}}_{\; \underline{\mathbf{u}}}$ can be expressed as: 
\begin{eqnarray}
\underline{\underline{\mathbf{C}}}_{\; \underline{\mathbf{u}}} =  \underline{\underline{\mathbf{G}}} \otimes_{T,S} \underline{\underline{\mathbf{C}}}_{\; \underline{\mathbf{f}}} \otimes_{T,S} \underline{\underline{\mathbf{G}}}^{-} 
\label{Cu2}
\end{eqnarray}
where $\underline{\underline{\mathbf{G}}}^{-} (t,\underline{x},\underline{x}'):=\underline{\underline{\mathbf{G}}}(-t,\underline{x},\underline{x}')^{T}$. It is important to note that to establish \textcolor{blue}{(\ref{Cu2})}, we use the property:  $\underline{\underline{\mathbf{G}}}(t,\underline{x},\underline{x}')=\underline{\underline{\mathbf{G}}}(t,\underline{x}',\underline{x})$,  true for all times $t$ and all couples of positions $(\underline{x},\underline{x}')$, according to spatial reciprocity. Equation \textcolor{blue}{(\ref{Cu2})} is the "order two" version of equation \textcolor{blue}{(\ref{u=Gf})}. When the source is a white noise, we obtain by using equation \textcolor{blue}{(\ref{Cu2})} an expression of the Green correlation: 
\begin{eqnarray}
\underline{\underline{\mathbf{C}}}:=\underline{\underline{\mathbf{G}}} \otimes_{T,S} \underline{\underline{\mathbf{G}}}^{-}  
\label{C}
\end{eqnarray}
This shows the fundamental importance of the Green correlation in medium identification when statistical properties of ambient sources are taking into account. Equations \textcolor{blue}{(\ref{Cu2})} and \textcolor{blue}{(\ref{C})} show that "perfect" white noise is for passive identification what "perfect" impulsion is for active identification.\\ 

	Green correlation does not appear in the literature as a fundamental field to retrieve in order to estimate medium parameters. Generally, the cross-correlation of a field generated by a white noise through a propagation medium is directly related to the Green function by the authors. This relation is called Ward identity and relates actually the Green correlation to the Green function.
	
\section{Elastic propagation with viscous damping.} 
\label{sec:3}

\indent \indent	In this section, we recall acoustic and solid propagation with viscous damping equations \textcolor{blue}{\cite{Royer}}. Classical results are derived with a vectorial formalism which is useful to describe solid waves propagation, in particular the coupling between the P-waves and the S-waves. Dispersion matrix and relation of dispersion are presented in order to compute the acoustic and solid Green function, the acoustic and solid Green correlation and Ward identities. Low attenuation case is also studied because it provides physical interpretations. 

\subsection{Solid waves equation with viscous damping.}
\label{sec:31}

\indent \indent	We consider an elastic, homogeneous, isotropic and linear solid medium. Let $(\lambda,\mu)$ be the Lam\'e parameters expressed in $N.m^{-2}$, $(\chi,\eta)$ be the viscous damping parameters for P-waves and S-waves, respectively, expressed in $N.s.m^{-2}$, and $\rho$ be the density of the medium expressed in $kg.m^{-3}$. Let $\underline{\mathbf{f}}$ be a 3-components causal spatio-temporal displacement source and $\underline{\mathbf{u}}$ the 3-components displacement field. From Newton theorem and Hooke law, we obtain the solid waves equation with viscous damping \textcolor{blue}{\cite{Royer}}:   
\begin{eqnarray}
\left[\frac{\partial^{2}}{\partial t^{2}} \mathcal{I}_{3} +  \frac{\partial}{\partial t}\mathcal{D} + \mathcal{L} \right]\underline{\mathbf{u}}=\underline{\mathbf{f}}
\label{NL}
\end{eqnarray} 
$\mathcal{I}_{3}$ is the identity operator of 3-components fields,
\begin{eqnarray}
 \mathcal{L}  :=  - \frac{\lambda+\mu}{\rho} \nabla \nabla ^{T} -  \frac{\mu}{\rho} \Delta \mathcal{I}_{3} & \mbox{ (Propagation operator)}  \nonumber \\
 \mathcal{D}  :=  - \frac{\chi+\eta}{\rho} \nabla  \nabla   ^{T} - \frac{\eta}{\rho} \Delta \mathcal{I}_{3} & \mbox{ (Dissipation operator)} \nonumber
\end{eqnarray} 
where $\Delta$ is the Laplacian operator and $\nabla$ is the gradient operator. \\

	According to the Helmholtz-Hodge theorem \textcolor{blue}{\cite{Arfken}}, each displacement field $\underline{\mathbf{u}}$ satisfying equation \textcolor{blue}{(\ref{NL})} when there is no source \textit{i.e.}  $\underline{\mathbf{f}}=0$, can be decomposed as $\underline{\mathbf{u}}=\nabla  \mathbf{\phi}_P+\nabla  \wedge \underline{\mathbf{\psi}}_S$, respectively, where $\wedge$ is the cross product. $\mathbf{\phi}_P$ is a scalar field, and, $\underline{\mathbf{\psi}}_S$ is a two-components field satisfying:
\begin{eqnarray}
\left[\frac{\partial^{2}}{\partial t^{2}} -  \alpha_P^2\frac{\partial}{\partial t}\Delta  - v_P^2 \Delta  \right] \mathbf{\phi}_P=0 \label{uP} \\
\left[\frac{\partial^{2}}{\partial t^{2}} \mathcal{I}_{2} -  \alpha_S^2\  \frac{\partial}{\partial t}\Delta -v_S^2 \Delta \right] \underline{\mathbf{\psi}}_S=0 \label{uS} 
\end{eqnarray} 	
with $v_P^2:=\frac{\lambda+2\mu}{\rho}$, $v_S^2:=\frac{\mu}{\rho}$, $\alpha_P^2:=\frac{\chi+2\eta}{\rho} $ and $\alpha_S^2:=\frac{\eta}{\rho}$. This shows that a solid wave is the contribution of two modes satisfying the classical acoustic waves equation with viscous damping \textcolor{blue}{\cite{Royer}}. $ \mathbf{\phi}_P$ is a pressure wave, called $P$-wave, of speed $v_P$ and attenuation $\alpha_P$. $ \underline{\mathbf{\psi}}_S$ is a shear waves, called $S$-wave, of speed $v_S$ and attenuation $\alpha_S$.

\subsection{Dispersion matrix and relation of dispersion.}
\label{sec:32}

\indent \indent	For a deterministic field $\underline{\mathbf{u}}$, we denoted by:
\begin{eqnarray}
\underline{\widehat{\mathbf{u}}}(\omega,\underline{k}):=\displaystyle{\int_{\mathbb{R} \times \mathbb{R}^3}} \underline{\mathbf{u}}(t,\underline{x})e^{-\mathbf{i} \left(\omega t - \underline{k}^{T}\underline{x} \right)} dtd\underline{x} 
\end{eqnarray}
its Fourier transform in the $(\omega,\underline{k})$-domain where $\omega$ and $\underline{k}$ are the frequency variables associated with $t$ and $\underline{x}$, respectively. \\
	
	Applying time and space Fourier transform to equation \textcolor{blue}{(\ref{NL})}, we obtain the algebraic relation: 
\begin{eqnarray}
\underline{\underline{\mathfrak{D}}}(\omega,\underline{k}) \underline{\hat{\mathbf{u}}}(\omega,\underline{k})  = \underline{\hat{\mathbf{f}}}(\omega,\underline{k}) 
\label{Du=f}
\end{eqnarray}
where $\underline{\underline{\mathfrak{D}}}(\omega,\underline{k})$ is the dispersion matrix: 
\begin{eqnarray}
\underline{\underline{\mathfrak{D}}}(\omega,\underline{k}) :=  (\textbf{v}_{P}^{2}(\omega)-\textbf{v}_{S}^{2}(\omega)) \underline{k} \, \underline{k}^{T} + (\textbf{v}_{S}^{2}(\omega)k^{2}-\omega^{2})\underline{\underline{\mathbf{I}_{3}}} \label{DM}
\end{eqnarray}
with: $\mathbf{v}_{P}^{2}(\omega) := v_{P}^{2}+\mathbf{i} \omega \alpha_{P}^{2}$, $
\mathbf{v}_{S}^{2}(\omega)  :=  v_{S}^{2}+\mathbf{i} \omega \alpha_{S}^{2} $, $k^{2}  :=  \underline{k}^{T} \, \underline{k} $ and $\underline{\underline{\mathbf{I}_3}}$ is the $3 \times 3$ identity matrix.  \\

	The dispersion matrix is fundamental as it plays the same role as the wave operator appearing in equation \textcolor{blue}{(\ref{NL})} but in the $(\omega,\underline{k})$-domain. Its decomposition into real and imaginary parts highlights roles of propagation operator $\mathcal{L}$ and dissipation operator $\mathcal{D}$, respectively, in that domain:
\begin{eqnarray}
\mathrm{Re} \, \underline{\underline{\mathfrak{D}}}(\omega,\underline{k}) &=& (v_{P}^{2}-v_{S}^{2}) \underline{k} \, \underline{k}^{T} + (v_{S}^{2}k^{2}-\omega^{2})\underline{\underline{\mathbf{I}_{3}}}
\label{ReD}\\
\mathrm{Im} \, \underline{\underline{\mathfrak{D}}}(\omega,\underline{k}) &=& \omega(\alpha_{P}^{2}-\alpha_{S}^{2}) \underline{k} \, \underline{k}^{T} + \omega \alpha_{S}^{2}k^{2} \underline{\underline{\mathbf{I}_{3}}} \label{ImD}
\end{eqnarray}
where $\mathrm{Re}$ and $\mathrm{Im}$ are the real and imaginary part operators. Then, the propagation is described by the real part of the dispersion matrix while the dissipation is described by its imaginary part.\\

	The relation of dispersion defined from the dispersion matrix by $\det \; \underline{\underline{\mathfrak{D}}}(\omega,\underline{k})=0$, where $\det$ is the determinant operator, gives the propagation modes. More precisely, by introducing the dispersion manifold $\mathbb{M}$ as the set of all couples $(\omega,\underline{k}) \in \mathbb{R} \times \mathbb{C}^{3}$ which satisfy the relation of dispersion, we show that $(\omega,\underline{k}) \in \mathbb{M}$ if and only if: 
\begin{eqnarray}
 k^{2} =  k^{2}_{P}(\omega) := \frac{\omega^{2}}{\mathbf{v}_{P}^{2}(\omega)} \mbox{ or } k^{2} =  k^{2}_{S}(\omega) := \frac{\omega^{2}}{\mathbf{v}_{S}^{2}(\omega)} 
 \label{disp}
\end{eqnarray} 
The dispersion manifold naturally appears as the union of two manifolds \textit{i.e.} $\mathbb{M} = \mathbb{M}_{P} \cup \mathbb{M}_{S}$ which corresponds to P-waves and S-waves with complex velocities $\mathbf{v}_{P}$ and $\mathbf{v}_{S}$, respectively. \\

	We consider now the low attenuation case. From the definition of $\mathbf{v}_P$ and $\mathbf{v}_S$ this assumption corresponds to $\omega \alpha_{P}^{2}/v_{P}^{2}$ $<<1$ and $\omega \alpha_{S}^{2}/v_{S}^{2}<<1$. Then, from \textcolor{blue}{(\ref{disp})} we get: 
\begin{eqnarray}
k_{P}(\omega) \approx \frac{\omega}{v_{P}} \; ; \; k_{S}(\omega) \approx   \frac{\omega}{v_{S}}
\label{dispa}
\end{eqnarray}
Those approximations of $k_{P}(\omega)$ and $k_{S}(\omega)$ will be useful to derive expression of the Green function and correlation and also Ward identities. 

\section{Elastic Green function.}
\label{sec:4}

\indent \indent	In this section, we introduce and compute in some representation domains the Green function of an elastic medium with viscous damping. We discuss the possibility to extract medium parameters from this field. The elastic Green function derived holds for unbounded medium;  however, it is still usable for bounded media, as long as different bouncing waves may be interpreted as coming from new sources. This will be valid if the boundaries are far enough from the points of interest. The same line of reasoning hold for dealing with potentially present weak heterogeneities. 

\subsection{Green function in the $(\omega,\underline{k})$-domain.}
\label{sec:41}	

\indent \indent	For an unbounded, isotropic, homogeneous solid media, the Green function depends only on the difference between its spatial parameters \textit{i.e.} we can do the following substitution: $\underline{\underline{\mathbf{G}}}(t,\underline{x},\underline{x}') \leftrightarrow \underline{\underline{\mathbf{G}}}(t,\underline{x}-\underline{x}')$.\\

	As the Green function is the response to an isotropic spatio-temporal impulsion source \textit{i.e} $\underline{\underline{\mathbf{f}}}:=\delta(t,\underline{x})\underline{\underline{\mathbf{I}_3}}$, we obtain from equation \textcolor{blue}{(\ref{Du=f})}: 
\begin{eqnarray}
\underline{\underline{\mathfrak{D}}}(\omega,\underline{k}) \underline{\underline{\hat{\mathbf{G}}}}(\omega,\underline{k})  = \underline{\underline{\mathbf{I}_{3}}}
\label{InvGok}
\end{eqnarray}
In that domain and according to equation \textcolor{blue}{(\ref{InvGok})}, the Green function appears to be the inverse of the dispersion matrix \textit{i.e.} $\underline{\underline{\hat{\mathbf{G}}}}(\omega,\underline{k})=\underline{\underline{\mathfrak{D}}}(\omega,\underline{k})^{-1}$. Using expression \textcolor{blue}{(\ref{DM})}, we show that:
\begin{eqnarray}
\underline{\underline{\hat{\mathbf{G}}}}(\omega,\underline{k})&=& \frac{\omega^{-2} k_{P}^{2}(\omega)}{k^{2}_{P}(\omega)-k^{2}}\tilde{\underline{k}} \, \tilde{\underline{k}}^{T}+\frac{\omega^{-2} k_{S}^{2}(\omega)}{k^{2}_{S}(\omega)-k^{2}}\left( \underline{\underline{\mathrm{I_{3}}}}- \tilde{\underline{k}} \, \tilde{\underline{k}}^{T}\right) \label{Gok} \\
&=:& \hat{\mathbf{G}}_{P}(\omega , \underline{k}) \tilde{\underline{k}} \, \tilde{\underline{k}}^{T} + \hat{\mathbf{G}}_{S}(\omega , \underline{k})\left( \underline{\underline{\mathrm{I_{3}}}}- \tilde{\underline{k}} \, \tilde{\underline{k}}^{T}\right)
\label{dGok}
\end{eqnarray}
where $\tilde{k} := \underline{k} /||\underline{k}||$ and $||.||$ is the classical Euclidean norm. The equation above highlights the decoupling between P-waves and S-waves in the $(\omega,\underline{k})$-domain. The P-waves displacement field, of amplitude determined  by $\hat{\mathbf{G}}_{P}(\omega , \underline{k})$, is on the axis directed by $\underline{k}$ whereas S-waves displacements, of amplitude determined  by $\hat{\mathbf{G}}_{S}(\omega , \underline{k})$, occur in the plane orthogonal to $\underline{k}$.\\

	It is important to note that $\hat{\mathbf{G}}_{P}(\omega , \underline{k})$ is the Green function of a propagation medium of acoustic waves with viscous damping. We recall that equation \textcolor{blue}{(\ref{uP})} describes the propagation in such media.

\subsection{Green function in the $(\omega,\underline{x})$-domain.}
\label{sec:42}	

\indent \indent	We compute now the solid Green function in the $(\omega,\underline{x})$-domain. In that domain, fields are capped by a $\vee$. \\

	Applying  inverse Fourier transform to equation \textcolor{blue}{(\ref{Gok})} with respect to $\underline{k}$, we show in \textcolor{blue}{appendix \ref{sec:81}} that:
\begin{eqnarray}
\underline{\underline{\check{\mathbf{G}}}}(\omega , \underline{x})  = & \frac{e^{-\mathbf{i} ||\underline{x}|| k_{P}(\omega)}}{4 \pi ||\underline{x}|| \mathbf{v}_{P}^{2}(\omega)}  \tilde{\underline{x}} \, \tilde{\underline{x}}^{T}  + \frac{e^{-\mathbf{i} ||\underline{x}|| k_{S}(\omega)}}{4 \pi ||\underline{x}|| \textbf{v}_{S}^{2}(\omega)}  \left( \underline{\underline{\mathrm{I_{3}}}}-\tilde{\underline{x}} \, \tilde{\underline{x}}^{T} \right) + O \left(\frac{1}{||\underline{x}||^{2}} \right) \label{Gox}  \left(\underline{\underline{\mathrm{I_{3}}}}-3\tilde{\underline{x}} \, \tilde{\underline{x}}^{T} \right) \\
   = &   \check{\mathbf{G}}_{P}(\omega , \underline{x})  \tilde{\underline{x}}\, \tilde{\underline{x}}^{T} + \check{\mathbf{G}}_{S}(\omega , \underline{x}) \left( \underline{\underline{\mathrm{I_{3}}}}-\tilde{\underline{x}}\, \tilde{\underline{x}}^{T} \right)+\check{\mathbf{G}}_{P,S}(\omega , \underline{x}) \left( \underline{\underline{\mathrm{I_{3}}}}-3\tilde{\underline{x}}\, \tilde{\underline{x}}^{T} \right) 
 \label{dGox} 
\end{eqnarray}
where $\tilde{\underline{x}}:=\underline{x}/||\underline{x}||$ and $O$ is the classical Landau notation for dominated functions. The complete expression of $\underline{\underline{\check{\mathbf{G}}}}$ is given in \textcolor{blue}{appendix \ref{sec:81}}. $\underline{\underline{\check{\mathbf{G}}}}$ can be decomposed into the sum of a near-field term $\check{\mathbf{G}}_{P,S} = O(||\underline{x}||^{-2}) $ and two far-field terms $\check{\mathbf{G}}_{P}$ and $\check{\mathbf{G}}_{S}= O(||\underline{x}||) $. A coupling term between the two waves dominates in the expression of the near-field  contribution:
\begin{eqnarray}
 \underline{\underline{\check{\mathbf{G}}}}(\omega , \underline{x})   \approx  \check{\mathbf{G}}_{P,S}(\omega , \underline{x}) \left( \mathrm{I_{3}}-3\tilde{\underline{x}} \, \tilde{\underline{x}}^{T} \right) 
 \end{eqnarray}
The far-field term highlights a decoupling between the two types of waves:
\begin{eqnarray}
\underline{\underline{\check{\mathbf{G}}}}(\omega , \underline{x})   \approx  \check{\mathbf{G}}_{P}(\omega , \underline{x})  \tilde{\underline{x}} \, \tilde{\underline{x}}^{T} + \check{\mathbf{G}}_{S}(\omega , \underline{x}) \left( \mathrm{I_{3}}-\tilde{\underline{x}} \,\tilde{\underline{x}}^{T} \right)
\end{eqnarray} 
We can note that we only retrieve the far-field contribution by taking the trace \textit{i.e.} $\mathrm{Tr}(\underline{\underline{\check{\mathbf{G}}}})=\check{\mathbf{G}}_{P}+2\check{\mathbf{G}}_{S}$. This property is true in all representation domains. \\

	When low attenuation case is considered, the Green function can then be approximated in the $(\omega,\underline{x})$-domain using \textcolor{blue}{(\ref{Gox})} and \textcolor{blue}{(\ref{dispa})} by: 
\begin{eqnarray}
\underline{\underline{\check{\mathbf{G}}}}(\omega,\underline{x})   &\approx&  \frac{e^{-\mathbf{i}\omega \frac{||\underline{x}||}{v_{P}} }}{4 \pi ||\underline{x}|| v_{P}^{2} }  \tilde{\underline{x}} \, \tilde{\underline{x}}^{T} + \frac{e^{-\mathbf{i}\omega \frac{||\underline{x}||}{v_{S}}} }{4 \pi ||\underline{x}||  v_{S}^{2}}  \left( \underline{\underline{\mathbf{I}_3}}-\tilde{\underline{x}} \, \tilde{\underline{x}}^{T} \right) \nonumber \\
 &+& \left[  \left(\dfrac{\mathbf{i}||\underline{x}|| \omega}{v_P}-1\right)e^{- \frac{\mathbf{i}\omega ||\underline{x}||}{v_{P}}}+\left(-\dfrac{\mathbf{i}||\underline{x}|| \omega}{v_S}+1\right) e^{-\frac{ \mathbf{i}\omega ||\underline{x}||}{v_{S}}} \right]\dfrac{\underline{\underline{\mathbf{I}_{3}}}-3\tilde{\underline{x}} \,\tilde{\underline{x}}^{T}}{{4 \pi ||\underline{x}||^{3} \omega^{2}}} 
\label{Gaox}
\end{eqnarray} 
Equation \textcolor{blue}{(\ref{Gaox})} is nothing but the classical solid Green function \textcolor{blue}{\cite{Aki}} in the $(\omega,\underline{x})$-domain for ideal case where no dissipation occurs. This approximation gives an easy interpretable expression of the Green function. Indeed, we observe pure phases in the far-field contribution for the two types of waves which provide information on $||\underline{x}||, v_P$ and $v_S$. This information can be completed with the near-field term. 

\subsection{Approximated Green function in the $(t,\underline{x})$-domain.}
\label{sec:43}	

\indent \indent	Computation of a general expression for $\underline{\underline{\mathbf{G}}}(t,\underline{x})$  from equation \textcolor{blue}{(\ref{Gox})} when dissipation occurs, is difficult. In the low attenuation case, we obtain: 
\begin{eqnarray}
\underline{\underline{\mathbf{G}}}(t,\underline{x}) \approx \frac{\delta (t-t_{P})}{4 \pi v_{P}^{2} ||\underline{x}||}\tilde{\underline{x}} \, \tilde{\underline{x}}^{T}+\frac{ \delta (t-t_{S}) }{4 \pi v_{S}^{2} ||\underline{x}||}\left(\underline{\underline{\mathrm{I_{3}}}} - \tilde{\underline{x}} \, \tilde{\underline{x}}^{T}\right) -\frac{t \; \Pi_{[t_{P} , t_{S}]}(t)}{4 \pi ||\underline{x}||^3}\left( \underline{\underline{\mathrm{I_{3}}}}-3\tilde{\underline{x}} \, \tilde{\underline{x}}^{T} \right)
\label{Gtx}
\end{eqnarray}
where $t_{P}:=||\underline{x}||/v_{P}$,  $t_{S}:=||\underline{x}||/v_{S}$ and $\Pi_{[t_{P} , t_{S}]}(t)=1$ if $t \in [t_{P} , t_{S}]$ and $\Pi_{[t_{P},  t_{S}]}(t)=0$ otherwise. \textcolor{blue}{(\ref{Gtx})} is the classical solid Green function \textcolor{blue}{\cite{Aki}} for ideal case in the $(t,\underline{x})$-domain. In this simplified situation, interpreting the far-field term is easy as we can observe two times of arrival corresponding to each type of waves. The particle displacement of P-waves occurs  along the  propagation axis and particle displacement of S-waves remains  in the orthogonal plane. For the near-field contribution, the coupling term controls the behavior of $\underline{\underline{\mathbf{G}}}$, and provides also information on $||\underline{x}||, v_P$ and $v_S$. 

\section{Elastic Green correlation and Ward identities.}
\label{sec:5}

\indent \indent In this section, we compute the solid Green correlation and we highlight its role in passive identification of solid media. Results are also expressed for the acoustic case and for the far-field and low attenuation case. Exact and approximated Ward identities are derived and compared to existing ones for a different damping model. 

\subsection{Computations in the $(\omega,\underline{k})$-domain.}
\label{sec:51}	

\indent \indent	As the Green function is space-shift invariant, the Green correlation also satisfies this property \textit{i.e.} we can do the following substitution: $\underline{\underline{\mathbf{C}}}(t,\underline{x},\underline{x}') \leftrightarrow \underline{\underline{\mathbf{C}}}(t,\underline{x}-\underline{x}')$.\\

	Using equation \textcolor{blue}{(\ref{C})}, we can compute the solid Green correlation in the $(\omega, \underline{k})$-domain. Indeed, applying the Fourier transform to equation \textcolor{blue}{(\ref{C})} and noting that $\widehat{\underline{\underline{\mathbf{G}}}^{-}}=\widehat{\underline{\underline{\mathbf{G}}}}^{\dagger}$ where $\dagger$ transpose and conjugate, we obtain: 
\begin{eqnarray}
\underline{\underline{\hat{\mathbf{C}}}}(\omega,\underline{k})   =   \underline{\underline{\hat{\mathbf{G}}}}(\omega,\underline{k})\underline{\underline{\hat{\mathbf{G}}}}(\omega,\underline{k})^{\dagger} 
\label{Cok} 
\end{eqnarray}	
The relation between the solid Green correlation and the solid Green function is purely algebraic in this domain.\\

	We can deduce a Ward identity in the $(\omega,\underline{k})$-domain using equation \textcolor{blue}{(\ref{Cok})} and applying the lemma established in \textcolor{blue}{appendix \ref{sec:82}} with $\underline{\underline{\mathbf{A}}}=\underline{\underline{\mathfrak{D}}}(\omega,\underline{k})=\underline{\underline{\hat{\mathbf{G}}}}(\omega,\underline{k})^{-1}$ which satisfies lemma assumptions because of the presence of dissipation:
\begin{eqnarray}
\omega \underline{\underline{\hat{\mathbf{C}}}}(\omega,\underline{k})   = -\underline{\underline{\hat{\mathbf{D}}}}(\underline{k})^{-1} \mathrm{Im} \, \underline{\underline{\hat{\mathbf{G}}}}(\omega,\underline{k})
\label{Wok}
\end{eqnarray}		
where $\underline{\underline{\hat{\mathbf{D}}}}(\underline{k}):=\omega^{-1} \mathrm{Im} \, \underline{\underline{\mathfrak{D}}}(\omega,\underline{k})$ satisfies:
\begin{eqnarray}
\underline{\underline{\hat{\mathbf{D}}}}(\underline{k})^{-1} = \alpha_P^{-2} k^{-2} \underline{\tilde{k}} \, \underline{\tilde{k}}^{T}+\alpha_S^{-2} k^{-2}  \left(\underline{\underline{\mathbf{I}_3}}-\underline{\tilde{k}}\underline{\tilde{k}}\right)
\label{InvDk}
\end{eqnarray}	
$\underline{\underline{\hat{\mathbf{D}}}}$ is the dispersion matrix of the dissipation operator $\mathcal{D}$. Ward identity \textcolor{blue}{(\ref{Wok})} shows that we can retrieve the imaginary part of the solid Green function from the solid Green correlation. In fact, by invoking Kramers-Kronïg theorem \textcolor{blue}{(\ref{Wok})} based on the time-causality of the Green function, we can retrieve the complete Green function from the Green correlation. The matricial proportionality term $\underline{\underline{\hat{\mathbf{D}}}}(\underline{k})^{-1}$ highlights the fundamental role of dissipation to establish \textcolor{blue}{(\ref{Wok})}. In fact, \textcolor{blue}{(\ref{Wok})} is true for every dissipation operator which is a partial differential operator. When the dissipation operator is considered as a constant operator as in \textcolor{blue}{\cite{Gouedard}}, \textcolor{blue}{\cite{ColinDeVerdiere}} and \textcolor{blue}{\cite{Lacoume}}, we retrieve that the Green correlation is directly proportional to the imaginary part of the Green function. In the solid case with viscous damping, the proportionally term \textcolor{blue}{(\ref{InvDk})} is matricial and each of its components is proportional to $k^{-2}$. \\

	We can now compute the solid Green correlation in the $(\omega,\underline{k})$-domain. Using Ward identity \textcolor{blue}{(\ref{Wok})} and expression \textcolor{blue}{(\ref{Gok})}, we get:	
\begin{eqnarray}
\underline{\underline{\hat{\mathbf{C}}}}(\omega,\underline{k})   &=&  \frac{\omega^{-4}|k_{P}(\omega)|^{4}}{|k_{P}^{2}(\omega)-k^{2}|^2}\tilde{\underline{k}} \, \tilde{\underline{k}}^{T} +\frac{\omega^{-4}|k_{S}(\omega)|^{4}}{|k_{S}^{2}(\omega)-k^{2}|^2 } \left( \underline{\underline{\mathbf{I}_{3}}}- \tilde{\underline{k}} \, \tilde{\underline{k}}^{T}\right)
\label{Cok2} \\
&=:& \hat{\mathbf{C}}_{P}(\omega , \underline{k}) \tilde{\underline{k}} \, \tilde{\underline{k}}^{T} + \hat{\mathbf{C}}_{S}(\omega , \underline{k})\left( \underline{\underline{\mathrm{I_{3}}}}- \tilde{\underline{k}} \, \tilde{\underline{k}}^{T}\right)
\label{dCok}
\end{eqnarray}
where $\hat{\mathbf{C}}_{P}(\omega , \underline{k})=|\hat{\mathbf{G}}_{P}(\omega , \underline{k})|^2$ and $\hat{\mathbf{C}}_{S}(\omega , \underline{k})=|\hat{\mathbf{G}}_{S}(\omega , \underline{k})|^2$.\\

	Expression \textcolor{blue}{(\ref{Cok2})} proves that retrieving the Green correlation is sufficient to estimate medium parameters: $v_P$, $v_S$, $\alpha_P$ and $\alpha_S$. More precisely, those parameters are contained in $k_{P}^{2}(\omega)$ and $k_{S}^{2}(\omega)$ which are the poles of $\underline{\underline{\hat{\mathbf{C}}}}(\omega,\underline{k})$. \\

	For P-waves, we obtain from \textcolor{blue}{(\ref{Wok})} a Ward identity applicable to acoustic waves with viscous damping:
\begin{eqnarray}
\omega \hat{\mathbf{C}}_P(\omega,\underline{k})   = -\dfrac{1}{\alpha_P^2 k^2} \mathrm{Im} \, \hat{\mathbf{G}}_P(\omega,\underline{k})
\label{Wokp}
\end{eqnarray}
In that domain, the acoustic Green correlation is proportional to the imaginary part of the acoustic Green function. According to equation, \textcolor{blue}{(\ref{Wokp})} dissipation appears to be fundamental even in this scalar case. The proportionally term $1/(\alpha_P^2k^2)$ is inverse of the dispersion matrix (of size $1 \times 1$) of the operator $\alpha_P^2\Delta$ which is the dissipation operator appearing in equation \textcolor{blue}{(\ref{uP})}.  	

\subsection{Computations in the $(\omega,\underline{x})$-domain.}
\label{sec:52}	
	
\indent \indent We compute now the Green correlation in the $(\omega,\underline{x})$-domain. Inverse Fourier transform of equation \textcolor{blue}{(\ref{Cok2})} with respect to $k$ gives (\textcolor{blue}{appendix \ref{sec:83}}): 
\begin{eqnarray}
\underline{\underline{\check{\mathbf{C}}}}(\omega,\underline{x})   = &  - \frac{\mathrm{Im} \left(e^{-\mathbf{i}||\underline{x}||k_{P}(\omega)} \right)}{4 \pi ||\underline{x}||\omega^{3} \alpha^{2}_{P}}  \tilde{\underline{x}}\tilde{\underline{x}}^{T} - \frac{\mathrm{Im} \left(e^{-\mathbf{i}||\underline{x}||k_{S}(\omega)} \right)}{4 \pi ||\underline{x}||\omega^{3} \alpha^{2}_{S} } \left( \underline{\underline{\mathbf{I}_{3}}}-\tilde{\underline{x}}\tilde{\underline{x}}^{T}\right) +
O \left(\frac{1}{||\underline{x}||^{2}} \right) \left(\underline{\underline{\mathbf{I}_3}}-3\tilde{\underline{x}}\tilde{\underline{x}}^{T} \right)\nonumber  \\
  = &   \check{\mathbf{C}}_{P}(\omega , \underline{x})  \tilde{\underline{x}}\tilde{\underline{x}}^{T} + \check{\mathbf{C}}_{S}(\omega , \underline{x}) \left( \underline{\underline{\mathbf{I}_{3}}}-\tilde{\underline{x}}\tilde{\underline{x}}^{T} \right)+\check{\mathbf{C}}_{P,S}(\omega , \underline{x}) \left( \underline{\underline{\mathbf{I}_{3}}}-3\tilde{\underline{x}}\tilde{\underline{x}}^{T} \right) 
\label{Cox}
\end{eqnarray}  
The complete expression of $\underline{\underline{\check{\mathbf{C}}}}$ is given in \textcolor{blue}{appendix \ref{sec:83}} and shows that we can extract all the physical parameters from \textcolor{blue}{(\ref{Cox})}. The decomposition above is similar to the decomposition into near-field and far-field of $\underline{\underline{\check{\mathbf{G}}}}$. As for the Green function, we can retrieve only the far-field contribution by taking the trace \textit{i.e.} $\mathrm{Tr}(\underline{\underline{\check{\mathbf{C}}}})=\check{\mathbf{C}}_{P}+2\check{\mathbf{C}}_{S}$. This property is also true in all Fourier domains. \\

	With far-field and low attenuation assumptions,  \textcolor{blue}{(\ref{Cox})} and \textcolor{blue}{(\ref{dispa})} leads to:
\begin{eqnarray}
\underline{\underline{\check{\mathbf{C}}}}(\omega,\underline{x})   \approx    \frac{\sin \left( \omega \frac{||\underline{x}||}{v_{P}} \right)}{4 \pi ||\underline{x}||\omega^{3} \alpha^{2}_{P}}  \tilde{\underline{x}} \, \tilde{\underline{x}}^{T} +    \frac{\sin \left( \omega \frac{||\underline{x}||}{v_{S}} \right) }{4 \pi ||\underline{x}||\omega^{3} \alpha^{2}_{S}}   \left( \underline{\underline{\mathbf{I}_{3}}}-\tilde{\underline{x}} \, \tilde{\underline{x}}^{T}\right)
\label{Caox}
\end{eqnarray} 
Relations \textcolor{blue}{(\ref{Cox})} and \textcolor{blue}{(\ref{Caox})} are fundamental as they prove that we can extract medium parameters from the Green correlation in the $(\omega,\underline{x})$-domain. This domain is accessible from the recorded field by an array of sensors using classical methods of discrete Fourier transform as the FFT (Fast Fourier Transform) algorithm. Green correlation expression \textcolor{blue}{(\ref{Caox})} shows that in far-field and low attenuation case, we can retrieve times of arrival $t_P=||\underline{x}||/v_P$ and $t_S=||\underline{x}||/v_S$ using a pseudo-pulsation estimator.  \\

	From Ward identity \textcolor{blue}{(\ref{Wok})}, we can derived an exact Ward identity in the $(\omega,\underline{x})$-domain:
\begin{eqnarray}
\omega \underline{\underline{\check{\mathbf{C}}}}(\omega,\underline{x})= -\mathcal{D}^{-1} \mathrm{Im} \, \underline{\underline{\check{\mathbf{G}}}}(\omega,\underline{x})
\label{Wox}
\end{eqnarray}	
where $\mathcal{D}^{-1}$ is the inverse of dissipation operator. This identity extends the classical one for acoustic waves in a sense that the Green correlation is "proportional" to the imaginary part of the Green function where the inverse of the dissipation appears in the proportionality term. However, in this solid case with viscous damping, the proportionality term is $\mathcal{D}^{-1}$ and it is not a constant operator. Then, \textcolor{blue}{(\ref{Wox})} is not sufficient to retrieve directly the Green function from the Green correlation. \\

	Similarly, we show that in the acoustic case, the Ward identity is:
\begin{eqnarray}
\omega \check{\mathbf{C}}_P(\omega,\underline{x})= -\alpha_P^{-2} \Delta^{-1} \mathrm{Im} \, \check{\mathbf{G}}_P(\omega,\underline{x})
\label{wox}
\end{eqnarray}
The "proportionality" term is the inverse of the Laplacian operator which is the dissipation operator. Even in this scalar case, Ward identity \textcolor{blue}{(\ref{wox})} is still hard to interpret. \\

	Using relations \textcolor{blue}{(\ref{Gaox})} and \textcolor{blue}{(\ref{Caox})}, we obtain the following approximated Ward identity: 
\begin{eqnarray}
\omega ^{3}\underline{\underline{\check{\mathbf{C}}}}(\omega,\underline{x}) &  \approx & - \underline{\underline{\mathbf{d}}}(\underline{x})^{-1} \; \mathrm{Im} \;  \underline{\underline{\check{\mathbf{G}}}}(\omega,\underline{x})
\label{Waox} 
\end{eqnarray} 
where: 
\begin{eqnarray}
 \underline{\underline{\mathbf{d}}}(\underline{x})^{-1} &  = &   \frac{v_{P}^{2}}{\alpha_{P}^{2}}   \tilde{\underline{x}} \, \tilde{\underline{x}}^{T} +  \frac{v_{S}^{2}}{\alpha_{S}^{2}} \left( \underline{\underline{\mathbf{I}_{3}}}-\tilde{\underline{x}}  \, \tilde{\underline{x}}^{T} \right) 
\end{eqnarray}
Identity \textcolor{blue}{(\ref{Waox})} highlights a matricial proportionality between the imaginary part of the approximated solid Green function and the approximated solid Green correlation. The proportionality term is $\omega^{-3} \underline{\underline{\mathbf{d}}}(\underline{x})^{-1}$ which form is interpretable using the dissipation term $\partial \mathcal{D}/\partial t $ in propagation equation \textcolor{blue}{(\ref{NL})} and relations of dispersion \textcolor{blue}{(\ref{dispa})}. Indeed, $\omega^{-1}$ is the contribution of operator $\partial/\partial t$ in the $(\omega,\underline{x})$-domain. $\omega^{-2}\underline{\underline{\mathbf{d}}}(\underline{x})^{-1}$ is the matricial contribution of $\mathcal{D}$, it has indeed a similar form as $\underline{\underline{\mathbf{D}}}(\underline{k})^{-1}$ \textcolor{blue}{(\ref{InvDk})} in the $(\omega,\underline{x})$-domain. The proportionality term $k^{-2}$ in \textcolor{blue}{(\ref{Wok})} has become the proportionality term $\omega^{-2}$ in \textcolor{blue}{(\ref{Waox})} due to the approximated relations of dispersion \textcolor{blue}{(\ref{dispa})} which link proportionality $k^2$ and $\omega^2$ for the two modes. \\

	Conserving only P-waves contribution in equation \textcolor{blue}{(\ref{Waox})}, it comes:
\begin{eqnarray}
\omega ^{3}\check{\mathbf{C}}_P(\omega,\underline{x})   \approx  - \frac{v_{P}^{2}}{\alpha_{P}^{2}}   \; \mathrm{Im} \;  \check{\mathbf{G}}_P(\omega,\underline{x}) 
\label{WaoxP} 
\end{eqnarray} 
In that domain, this approximated Ward identity highlights a proportionality between acoustic Green function and correlation. 
	
\subsection{Computations in the $(t,\underline{x})$-domain.}
\label{sec:53}	

\indent \indent	 The solid Green correlation in the $(t,\underline{x})$-domain can be obtained  by Fourier transforming equation \textcolor{blue}{(\ref{Cox})}. The complicated analytical form of the exponential terms makes this calculation difficult. However, we can deduce from Ward identity \textcolor{blue}{(\ref{Wox})}, an exact Ward identity in the $(t,\underline{x})$-domain:
\begin{eqnarray}
\dfrac{\partial \underline{\underline{\mathbf{C}}}(t,\underline{x})}{\partial t}   = -\mathcal{D}^{-1} \mathrm{Odd} \, \underline{\underline{\mathbf{G}}}(t,\underline{x})
\label{Wtx}
\end{eqnarray}	
where $\mathrm{Odd} \, \underline{\underline{\mathbf{G}}}:=1/2 \left(\underline{\underline{\mathbf{G}}}-\underline{\underline{\mathbf{G}}}^-\right)$ is the odd part of the Green function. In the acoustic case, Ward identity \textcolor{blue}{(\ref{wox})} leads to: 
\begin{eqnarray}
\dfrac{\partial \mathbf{C}_P(t,\underline{x})}{\partial t}   = -\alpha_P^{-2} \Delta^{-1}  \mathrm{Odd} \, \mathbf{G}_P(t,\underline{x})
\label{wtx}
\end{eqnarray}	
As identities \textcolor{blue}{(\ref{Wox})} and \textcolor{blue}{(\ref{wox})}, identities \textcolor{blue}{(\ref{Wtx})} and \textcolor{blue}{(\ref{wtx})} extend classical acoustic results \textcolor{blue}{\cite{Gouedard}}, \textcolor{blue}{\cite{ColinDeVerdiere}}, \textcolor{blue}{\cite{Lacoume}} with constant dissipation, but are difficult to apply in practice because of the presence of the inverse of dissipation operator $\mathcal{D}$ is the solid case and operator $\alpha_P^2\Delta$ in the acoustic case.\\ 

	We can get over this last step in the far-field and low attenuation situation. From equation \textcolor{blue}{(\ref{Waox})}, we obtain: 
\begin{eqnarray}
\frac{\partial^{3} \underline{\underline{\mathbf{C}}}(t,\underline{x})}{\partial t^{3}}   \approx  -  \underline{\underline{\mathbf{d}}}(\underline{x})^{-1}  \; \mathrm{Odd} \;  \underline{\underline{\mathbf{G}}} (t,\underline{x}) 
\label{Watx}
\end{eqnarray}  
The presence of the third-time derivative is justified by the presence of the term $\omega^3$ in identity \textcolor{blue}{(\ref{Waox})}. Finally, in the solid case with viscous damping, when far-field propagation and low attenuation are considered, the third-time derivative of the solid Green correlation is matricially proportional to the odd part of the solid Green function. The proportionally term is due to the dissipation and \textcolor{blue}{(\ref{Waox})} highlights its fundamental role to establish a Ward identity in that domain. \\

	Retaining only P-waves terms in approximated Ward identity \textcolor{blue}{(\ref{Watx})}, we obtain:
\begin{eqnarray}
\frac{\partial^{3} \mathbf{C}_P(t,\underline{x})}{\partial t^{3}}   \approx  -  \frac{v_{P}^{2}}{\alpha_{P}^{2}}    \; \mathrm{Odd} \;  \mathbf{G}_P (t,\underline{x}) 
\label{watx}
\end{eqnarray} 
Again, we can easily make an analogy with the classical identities obtained in the acoustic framework when the dissipation is supposed to be constant \textcolor{blue}{\cite{Gouedard}}, \textcolor{blue}{\cite{ColinDeVerdiere}}, \textcolor{blue}{\cite{Lacoume}}. In that case and domain, it is shown that the first-time derivative of the Green correlation is proportional to the odd part of the Green function with a proportionality term inversely proportional to the constant dissipation. For a viscous damping and for far-field and low attenuation framework, it is the third-time derivative which is directly proportional to the odd part of the Green function in the $(t,\underline{x})$-domain. The difference derivative order comes only from the form of the dissipation.
	
\section{Conclusion.}
\label{sec:7}

\indent \indent Our contributions can be summarized into two points. Firstly, we introduced the Green correlation in a general context. We presented its fundamental role in passive identification in the sense that all medium parameters can be retrieved from this field and without retrieving the Green function from a Ward identity. \\
\indent Secondly, we applied this theory to acoustic and solid waves with viscous damping. We computed, in different representation domains, the Green function, the Green correlation and associated Ward identities. Those latter highlight the fundamental role played by the dissipation operator as its frequency law describes the proportionality between the elastic Green function and correlation. Equations \textcolor{blue}{(\ref{Cox})} and \textcolor{blue}{(\ref{Caox})} give Green correlation expressions in an accessible domain from field measurements. Those expressions provides all mechanical information on the medium which highlights  Green correlation approach in passive identification of elastic media. We considered the far-field and low attenuation case, which provides interesting and interpretable Ward identities in $(\omega,\underline{x})$- and $(t,\underline{x})$-domains. More precisely, in that case, it is the third time-derivative of the Green correlation which is proportional to the odd part of the Green function \textcolor{blue}{(\ref{Watx})}, \textcolor{blue}{(\ref{watx})}. This is a direct consequence of viscous damping model and this result was compared to classical ones in acoustic with other damping models.\\

	The perspective pursued is to apply experimentally our results to solid media with an embedded 3-components instrumentation in order to retrieve the complete solid Green function and correlation \textit{i.e.} the whole matrices. It is a very interesting perspective as it has never been realised, from our knowledge. Role of sensor orientations will certainly be emphasized. 

\section{Appendix.}
\label{sec:8}

\indent \indent All non direct calculus done to derive the Green function, the Green correlation and Ward identity are presented in this appendix. 

\subsection{Computations for the elastic Green function.}
\label{sec:81}

\indent \indent First, we need the following lemma: \\

\noindent \textit{\textbf{Lemma.} For $a \in \mathbb{C} \setminus \mathbb{R}$, let $Y(a,.) : \underline{k} \in \mathbb{R}^{3} \mapsto \frac{1}{k^{2}-a^2}$ and $\check{Y}(a,.)$ be its inverse Fourier transform. We can show that: 
\begin{eqnarray*}
\check{Y}(a,\underline{x}) &:=& \frac{1}{(2 \pi)^{3}} \int_{\mathbb{R}^{3}} Y \left( a,\underline{k} \right) e^{\mathbf{i}\underline{k}^T\underline{x}} d\underline{k}=\frac{e^{-\mathbf{i} a ||\underline{x}||}}{4 \pi ||\underline{x}||} \\
\bigtriangledown \bigtriangledown^{T} \check{Y}(a,\underline{x})& = &\check{Y}(a,\underline{x})\left[ -a^{2} \tilde{\underline{x}} \, \tilde{\underline{x}}^{T}+\frac{\mathbf{i}a||\underline{x}||-1}{||\underline{x}||^{2}}\left( \underline{\underline{\mathbf{I}_{3}}}-3\tilde{\underline{x}} \, \tilde{\underline{x}}^{T}\right)  \right]  
\end{eqnarray*}
Then $a \mapsto \check{Y}(a,)$ can be prolonged to whole of $\mathbb{C}$.}\\

	The starting point is \textcolor{blue}{(\ref{Gok})}, which can be rewritten as: 
\begin{eqnarray}
\underline{\underline{\hat{\mathbf{G}}}}(\omega,\underline{k})=- \omega^{-2} Y(k_{P}(\omega),\underline{k}) \underline{k} \, \underline{k}^{T} - \omega^{-2}Y(k_{S}(\omega),\underline{k}) \left(k_{S}^{2}(\omega)\underline{\underline{\mathbf{I}_{3}}}-  \underline{k} \, \underline{k}^{T} \right)
\label{A1}
\end{eqnarray}
Using the lemma, we have: 
\begin{eqnarray*}
\underline{\underline{\check{\mathbf{G}}}}(\omega,\underline{x}) & = & \omega^{-2}  \bigtriangledown \bigtriangledown ^{T} \check{Y} \left( k_{P}(\omega),\underline{x} \right) + \omega^{-2} \left(k_{S}^{2}(\omega)\mathcal{I}_3-  \bigtriangledown \bigtriangledown ^{T} \right)\check{Y} \left( k_{S}(\omega),\underline{x} \right) \\
& = & \frac{k_{P}^{2}(\omega)e^{-\mathbf{i} ||\underline{x}|| k_{P}(\omega)}}{4 \pi ||\underline{x}|| \omega^{2}}  \tilde{\underline{x}} \, \tilde{\underline{x}}^{T} +\frac{k_{S}^{2}(\omega)e^{-\mathbf{i} ||\underline{x}|| k_{S}(\omega)}}{4 \pi ||\underline{x}|| \omega^{2}}  \left( \underline{\underline{\mathbf{I}_{3}}}-\tilde{\underline{x}} \, \tilde{\underline{x}}^{T} \right)+ \\
 &&  \left[  \left(\mathbf{i}||\underline{x}||k_{P}(\omega)-1\right)e^{-\mathbf{i} ||\underline{x}|| k_{P}(\omega)}+\left(-\mathbf{i}||\underline{x}||k_{S}(\omega)+1\right) e^{-\mathbf{i} ||\underline{x}|| k_{S}(\omega)} \right]\dfrac{\underline{\underline{\mathbf{I}_{3}}}-3\tilde{\underline{x}} \,\tilde{\underline{x}}^{T}}{{4 \pi ||\underline{x}||^{3} \omega^{2}}} 
\end{eqnarray*}
This equation gives the complete and exact expression of elastic Green function with viscous damping in the $(\omega,\underline{x})$-domain. 

\subsection{Fundamental lemma for Ward identity.}
\label{sec:82}

We consider the following lemma:\\

\noindent \textit{\textbf{Lemma.} Let $\underline{\underline{\mathbf{A}}}$ be an invertible matrix such that $\mathrm{Im} \, \underline{\underline{\mathbf{A}}}:=
1/2(\underline{\underline{\mathbf{A}}}-\underline{\underline{\mathbf{A}}}^{\dagger})$ is invertible and $\underline{\underline{\mathbf{A}}} \,\underline{\underline{\mathbf{A}}}^{\dagger}=
\underline{\underline{\mathbf{A}}}^{\dagger}\underline{\underline{\mathbf{A}}}$. Then we have:  $\underline{\underline{\mathbf{A}}}^{-1}\underline{\underline{\mathbf{A}}}^{{\dagger} \,-1}=-\left(\mathrm{Im} \, \underline{\underline{\mathbf{A}}} \right)^{-1} \, \mathrm{Im} \left(\underline{\underline{\mathbf{A}}}^{-1} \right) $.}\\

\noindent \textit{Proof.} The lemma is a direct consequence of the equality: $\left(\mathrm{Im} \, \underline{\underline{\mathbf{A}}} \right)\underline{\underline{\mathbf{A}}}^{-1}\underline{\underline{\mathbf{A}}}^{{\dagger} \,-1}=-\mathrm{Im} \left(\underline{\underline{\mathbf{A}}}^{-1} \right)$. This equality can be proved by expanding the left hand side using the definition of $\mathrm{Im} \, \underline{\underline{\mathbf{A}}}$.  

\subsection{Computations for the elastic Green correlation.}
\label{sec:83}

\indent \indent Using equations (\ref{Cok}) and (\ref{A1}), we have: 
\begin{eqnarray*}
\underline{\underline{\hat{\mathbf{C}}}}(\omega,\underline{k}) =  - \dfrac{\mathrm{Im} \, \left[k_{P}^{-2}(\omega) Y \left( k_{P}(\omega),\underline{k} \right) \underline{k} \, \underline{k}^{T}\right]}{\omega^{3} \alpha^{2}_{P} } -\dfrac{ \mathrm{Im} \, \left[ Y \left( k_{S}(\omega),\underline{k} \right)\left(\underline{\underline{\mathbf{I}_{3}}}-k_{S}^{-2}(\omega)\underline{k} \, \underline{k}^{T}\right)\right]}{\omega^{3} \alpha^{2}_{S} }
\end{eqnarray*}
then,
\begin{eqnarray*}
 \underline{\underline{\check{\mathbf{C}}}}(\omega,\underline{x}) &  = &  \dfrac{\mathrm{Im} \, \left[k_{P}^{-2}(\omega) \bigtriangledown \bigtriangledown ^{T} \check{Y} \left( k_{P}(\omega),\underline{x} \right)\right]}{\omega^{3} \alpha^{2}_{P} }-\dfrac{ \mathrm{Im} \, \left[\left(\mathcal{I}_{3}-k_{S}^{-2}(\omega)\bigtriangledown \bigtriangledown ^{T}\right) \check{Y} \left( k_{S}(\omega),\underline{x} \right)\right]}{\omega^{3} \alpha^{2}_{S} } \\
& = &  - \frac{\mathrm{Im} \, \left(e^{-\mathbf{i}||\underline{x}||k_{P}(\omega)} \right)}{4 \pi ||\underline{x}||\omega^{3} \alpha^{2}_{P}}  \tilde{\underline{x}} \, \tilde{\underline{x}}^{T} - \frac{\mathrm{Im} \, \left(e^{-\mathbf{i}||\underline{x}||k_{S}(\omega)} \right)}{4 \pi ||\underline{x}||\omega^{3} \alpha^{2}_{S} } \left( \underline{\underline{ \mathbf{I}_{3}}}-\tilde{\underline{x}} \, \tilde{\underline{x}}^{T}\right)+\\
 & &  \mathrm{Im} \, \left[ \frac{\mathbf{i} ||\underline{x}||k_{P}(\omega)-1}{\alpha^{2}_{P} k_{P}^{2}(\omega)}e^{-\mathbf{i}||\underline{x}||k_{P}(\omega)} +\frac{-\mathbf{i} ||\underline{x}||k_{S}(\omega)+1}{\alpha^{2}_{S}k_{S}^{2}(\omega)}e^{-\mathbf{i}||\underline{x}||k_{S}(\omega)} \right]  \dfrac{\underline{\underline{\mathbf{I}_{3}}}-3\tilde{\underline{x}} \,\tilde{\underline{x}}^{T}}{4 \pi ||\underline{x}||^{3}\omega^{3}}
\end{eqnarray*}
This is the complete and exact expression of the elastic Green correlation with viscous damping in the $(\omega,\underline{x})$-domain.


\begin{thebibliography}{}

\bibitem{Aki} K. Aki, P.G. Richards, \textit{Quantitative seismology}, University Science Books, 2002. 
\bibitem{Arfken} G.B. Arfken, H.J. Weber, \textit{Mathematical Methods for Physicists}, Elsevier, 6th edition, 2005.
\bibitem{Campillo} M. Campillo, A. Paul \textit{Long-Range Correlations in the diffuse seismic coda}, Science \textbf{299}, 2003, p. 547-549. 
\bibitem{Campillo2} M. Campillo, \textit{Phase and Correlation in "Random" Seismic Fields and
the Reconstruction of the Green Function}, Pure and Applied Geophysics \textbf{163}, 2006, p. 475-502.
\bibitem{ColinDeVerdiere} Y. Colin de Verdiere, \textit{Semi-classical analysis and passive imaging}, NonLinearity \textbf{22}, 2009, R45-R75.
\bibitem{Gouedard} P. Gouedard \textit{and al.}, \textit{Cross-correlation of random fields: mathematical approach and applications}, Geophysical Prospecting \textbf{56}, 2008, p. 375-393.
\bibitem{Lacoume} J.L. Lacoume, \textit{Tomographie passive: observer avec du bruit}, Colloque GRETSI, 2007. 
\bibitem{Landau} L. Landau, E. Lifchitz, A. Kosevich \textit{Physique théorique : théorie de l'élasticité}, 2nd edition, Editions MIR, 1990. 
\bibitem{Ljung} L. Ljung, \textit{System Identification: Theory for the User}, Prentice Hall, 2nd edition, 1999. 
\bibitem{Lobkis} O. Lobkis, R. Weaver, \textit{On the emergence of the Green function in the correlations of a diffuse field}, JASA \textbf{110}, 2001, p. 3011-3017.
\bibitem{Miyazawa} M. Miyazawa, R. Snieder, A. Venkataraman, \textit{Application of seismic interferometry to extract P- and S-wave propagation and observation of shear-wave splitting from noise data at Cold Lake, Alberta, Canada}, Geophysics \textbf{73}, 2008, p. 35-40.
\bibitem{Royer} D. Royer, E. Dieulesaint, \textit{Elastic Waves in Solids}, Springer, 2000. 
\bibitem{Sabra} K.G. Sabra \textit{and al.}, \textit{Extracting time-domain Green function estimates from ambient seismic noise}, Journal of Ocean Engineering \textbf{30}, 2005, p. 338-347. 
\bibitem{Snieder} R. Snieder, K. Wapenaar, U. Wegler, \textit{Unified Green function retrieval by cross-correlation; connection with energy principles}, Physical Review E \textbf{75}, 2008, 14 p.
\bibitem{Wapenaar} K. Wapenaar, E. Slob, R. Snieder, and A. Curtis, 2010, \textit{Tutorial on seismic interferometry. Part II: Underlying theory and new advances}, Geophysics \textbf{75}, 2010, \textit{in press}.
\bibitem{Weaver} R. Weaver, \textit{Ward identities and the retrieval of Green's functions in the correlations of a diffuse field},Wave Motion \textbf{45}, 2008, p. 596-604.
\end{thebibliography}
\end{document}